\documentclass{article}
\usepackage{spconf}
\usepackage{epsfig}
\usepackage{amsmath}
\usepackage{amsfonts}
\usepackage{amssymb}
\usepackage{bm}
\usepackage{nicefrac}
\usepackage{subcaption}
\usepackage{hyperref}
\usepackage{xcolor}
\usepackage{fancyhdr}

\let\OLDthebibliography\thebibliography
\renewcommand\thebibliography[1]{
  \OLDthebibliography{#1}
  \small
  \setlength{\parskip}{0pt}
  \setlength{\itemsep}{0pt plus 0.3ex}
}

\fancypagestyle{copyright}{
    \fancyhf{}
    \fancyhead[C]{
        \small
        \textit{IEEE ICME Workshop on Coding for Machines, Brisbane, Australia, 2023.} \\
        \vspace{10pt}
        © 2023 IEEE. Personal use of this material is permitted. Permission from IEEE must be obtained for all other uses, in any current or future media, including reprinting/republishing this material for advertising or promotional purposes, creating new collective works, for resale or redistribution to servers or lists, or reuse of any copyrighted component of this work in other works.
    }
    
}

\begin{document}
\title{Conditional and Residual Methods in Scalable Coding for Humans and Machines}
%
\name{Anderson de Andrade, Alon Harell, Yalda Foroutan, and Ivan V. Bajić}
\address{
School of Engineering Science, Simon Fraser University, Burnaby, Canada \\
\normalsize{
\texttt{\{anderson\_de\_andrade,alon\_harell,yalda\_foroutan\}@sfu.ca}, \texttt{ibajic@ensc.sfu.ca}
}}

\maketitle
\thispagestyle{copyright}

\begin{abstract}
We present methods for conditional and residual coding in the context of \textit{scalable coding for humans and machines}. Our focus is on optimizing the rate-distortion performance of the reconstruction task using the information available in the computer vision task. We include an information analysis of both approaches to provide baselines and also propose an entropy model suitable for conditional coding with increased modelling capacity and similar tractability as previous work. We apply these methods to image reconstruction, using, in one instance, representations created for semantic segmentation on the Cityscapes dataset, and in another instance, representations created for object detection on the COCO dataset. In both experiments, we obtain similar performance between the conditional and residual methods, with the resulting rate-distortion curves contained within our baselines.
\end{abstract}
\section{Introduction}
\label{sec:intro}

With the prominence of artificial intelligence, digital content is not only consumed by humans but also by computer programs. This software often analyzes content in different ways, according to their purpose. Depending on their task, only a subset of the information available might be necessary. Moreover, the information required can be represented in a more suitable way for the computer program that does not necessarily resemble its original natural representation, often required by humans to consume such content.

In a collaborative setting \cite{bajic2021collaborative}, where edge devices capture signals that are processed and transmitted to cloud 
services to complete a 
set of tasks, it is efficient to transmit only the information necessary to achieve these tasks. Creating representations for every subset of tasks does not scale well with the number of tasks. In addition, if information for some tasks has already been transmitted and a superset of the original tasks is now required for the same input, transmitting the new corresponding representation would incur 
an overhead in redundant information. Thus, we would like to compose the information required for tasks in a scalable fashion \cite{SchwarzMW07}, in which base representations are shared among multiple tasks and only incremental amounts of information are required for more specific tasks. 

Creating learnable tasks that make use of different streams of information in which some are fitted for different purposes is a challenge \cite{Choi2022ScalableIC}. The lower-dimensional manifold induced by a particular task might not be readily usable by a different task. Translating representations from one manifold to another, such that the maximum amount of information is usable in a secondary task, is limited by the modelling capacity of the transformation and the data available \cite{CulpepperO09, ConnorCR21}. Conditional and residual coding have prevailed as two different approaches to incorporate side information in learnable compression settings. These approaches can leverage dedicated learnable transformations to explicitly transfer information to the target domain.

We limit our findings to a common setting in which we have an image reconstruction task and a computer vision task whose representation is shared with the former. This configuration is referred as \textit{scalable image coding for humans and machines} \cite{Choi2022ScalableIC}. We present conditional and residual approaches for scalable learnable compression in which we transform the representations to share a common feature space. We derive baselines for these approaches and empirically compare them. Our experiments perform reconstruction of different datasets using representations for semantic image segmentation and object detection. We also present an entropy model with increased modelling potential suitable for conditional coding.

\section{Related Work}
\begin{figure*}[t]
  \centering
  \begin{subfigure}{.44\textwidth}
      \centering
      \includegraphics[width=\textwidth]{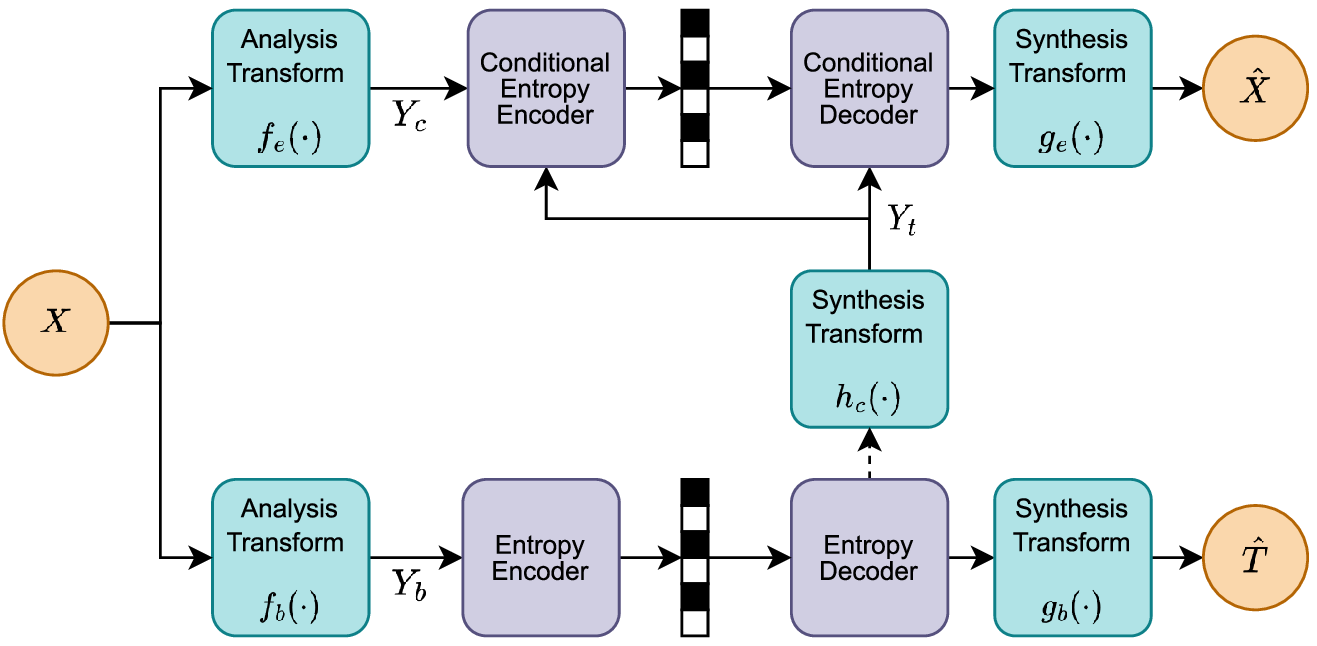}
      \caption{Conditional}
  \end{subfigure}
  \begin{subfigure}{.51\textwidth}
      \centering
      \includegraphics[width=\textwidth]{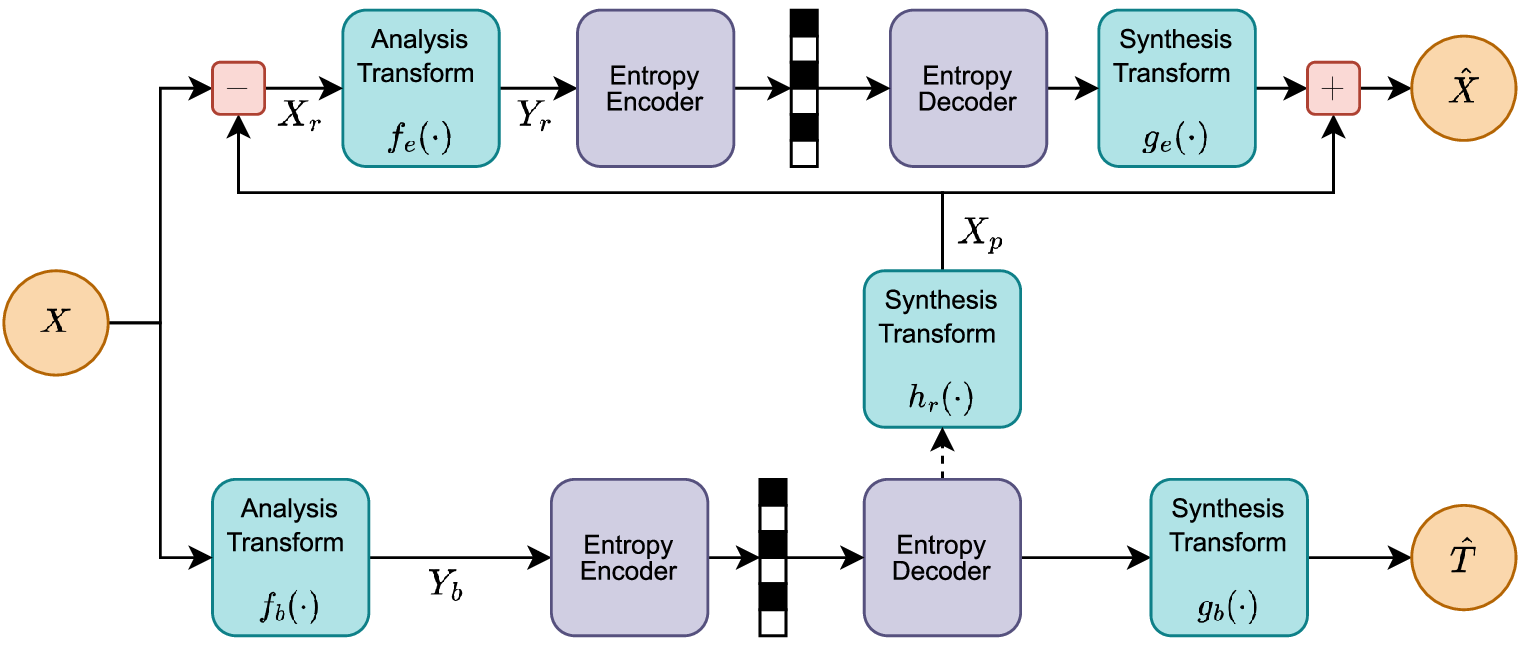}
      \caption{Residual}
  \end{subfigure}
  \caption{Overall architecture of the residual and conditional methods. The dotted line signifies that the enhancement network does not affect the base network. The conditional entropy decoder models $H(Y_c|Y_t)$.}
  \label{figure:architecture}
\end{figure*}

In learnable compression, an information bottleneck is induced on an intermediate representation between the input and the output \cite{tishby2000information}. Successful approaches follow a variational framework \cite{Kingma2014AutoEncodingVB} in which a hyper-prior representation learns the dependencies between the different factors of a latent representation and operates as side information \cite{balle2018variational, Minnen2018JointAA, Cheng2020LearnedIC, HeYPMQW22}. An auto-regressive entropy model is used to induce the information bottleneck and to entropy code the 
learnt representations.

Recent work on scalable coding for humans and machines apply the ideas of learnable compression to both the reconstruction and computer vision tasks \cite{ChoiB21, Choi2022ScalableIC, ozyilkan2023learned}. In these approaches, the reconstruction task uses a dedicated and a shared representation. These representations are concatenated after being decoded, and are used as input for a reconstruction model. Through rate-distortion optimization, this approach could create independent representations with no much redundancy of information between them, but the results of \cite{Choi2022ScalableIC} and \cite{ozyilkan2023learned} show considerable redundancy. This work focuses primarily on efficiently coding the dedicated representation for the reconstruction task. In the conditional approach, the dedicated representation contains all information relevant to reconstruction but the uncertainty resolved by the shared representation is exploited during coding. In the residual approach, the information of the shared representation is removed from the target representation before coding it, and then added back down the pipeline after decoding.

Residual and conditional coding in the context of learnable compression has been explored before for video compression \cite{DjelouahCSS19, ladune2021conditional, BrandSK22, hadizadeh2022lccmvc}. Our formulation is different in that the prediction is completely explained by the original signal and as such, the information of the residual cannot increase, whereas in video compression this can occur since the previous frame is used to compute the prediction. In \cite{BrandSK22}, it is shown that learnable conditional coding often requires a transformation of the side information, potentially resulting in information loss to a degree where a residual approach could outperform the conditional approach. In this work, we propose to transform the side information in both approaches and show how this can improve the performance of the residual approach with respect to the conditional approach. 

Many of the existing entropy models for learnable compression support the conditional coding of the target representation given the hyper-prior representation \cite{balle2018variational}. Recent entropy models extend these ideas to efficiently capture the spatial and dimensional dependencies of these representations, by grouping factors together \cite{MinnenS20}, and reorganizing and parallelizing the decoding order of the spatial locations \cite{HeZSWQ21}. In this work, we utilize our conditional information in a similar fashion, but we increase the modelling capacity by augmenting its receptive field and adding scaled residual connections.

\section{Proposed Methods}

For an input image $X \in \mathbb{R}^{C_x \times H \times W}$, a lossily-compressed \textit{base} representation $Y_b = f_b(X)$ is learned as to minimize the distortion $D_b = \mathbb{E}_X[d_b(g_b(Y_b), T)]$ with respect to a given computer vision target $T$, a task distortion function $d_b(\cdot, \cdot)$, and a learnable decoding function $g_b(\cdot)$.

In the conditional setting, a lossily-compressed \textit{enhancement} representation $Y_c = f_e(X)$ is learned to minimize the distortion $D_c = \mathbb{E}[d_e(\widehat{X}, X)]; \widehat{X} = g_e(Y_c)$, using an image reconstruction distortion function $d_e(\cdot, \cdot)$ and a learnable decoder $g_e(\cdot)$. All information used for the reconstruction task is contained in $Y_c$, and the information contained in $Y_b$ is utilized to efficiently code $Y_c$. Conditional coding effectively models $H(Y_c|Y_t)$, where $Y_t = h_c(Y_b)$ is a learnable transformation of $Y_b$ that intuitively has a feature space similar to that of the enhancement representation $Y_c$, so that their similarities can be exploited. Any information that reduces the conditional entropy should be maintained in $Y_t$ since there is no rate penalty on its rate.

In the residual approach, an analogous representation $Y_r = f_e(X_r); X_r = X - X_p; X_p = h_r(Y_b)$ is created to minimize $D_r = \mathbb{E}[d_e(g_e(Y_r) + X_p, X)]$. Here, $h_r(\cdot)$ is a learnable transformation of $Y_b$ that implicitly reconstructs the image. The prediction $X_p$ is added at the end of the reconstruction process.  Fig.~\ref{figure:architecture} shows architecture diagrams for both configurations.\footnote{Official code release: \href{https://github.com/adeandrade/research}{https://github.com/adeandrade/research}}

\subsection{Bounds for conditional coding}
\label{section:conditional-bounds}

Our theoretical analysis is performed in the  lossless case to motivate the proposed baselines for our lossy approaches. In conditional coding, we model $H(Y_b) + H(Y_c|Y_t)$, having $H(Y_c)$ as a lower bound:
\begin{align}
    \label{equation:cond-lower-bound}
     H(Y_c)
     &\leq H(Y_c) + H(Y_t|Y_c)
     = H(Y_c,Y_t) \nonumber \\
     &= H(Y_t) + H(Y_c | Y_t)
     \leq H(Y_b) + H(Y_c|Y_t)
     ,
\end{align}
where we used $H(Y_t) \leq H(Y_b)$ due to the data processing inequality. This bound is tight when $H(Y_t|Y_c) = 0$ and $H(Y_b|Y_t) = 0$, or equivalently, when $H(Y_b) = I(Y_c;Y_t)$. This corresponds to a decrease of information in $H(Y_c|Y_t)$ of $H(Y_b)$. An upper bound is obtained by:
\begin{align}
    \label{equation:cond-upper-bound}
    H(Y_b) + H(Y_c|Y_t)
    &= H(Y_b) + H(Y_c) - I(Y_c;Y_t) \nonumber \\
    &\leq H(Y_b) + H(Y_c)
    .
\end{align}
This bound is tight when $I(Y_c;Y_t) = 0$, which corresponds to 
$Y_c$ and $Y_t$ being independent.

We provide an upper baseline for the conditional approach by using a standalone enhancement representation $Y_e$ generated without relying on any side information, and measuring $\hat{H}(Y_e)$, where $\hat{H}(\cdot)$ is an entropy estimate. As a lower baseline we use $\hat{H}(Y_b) + \hat{H}(Y_e)$. This is motivated by considering that $Y_e$ can be more efficient as a task representation than $Y_c$, and by the bounds in \eqref{equation:cond-lower-bound} and \eqref{equation:cond-upper-bound}. 

\subsection{Bounds for residual coding}
It has been shown that conditional coding is an upper bound of residual coding \cite{BrandSK22, hadizadeh2022lccmvc}:
\begin{align}
    H(X | X_p)
    &= H(X_r + X_p | X_p)
    = H(X_r | X_p)
    \label{equation:summands} \\
    &= H(X_r) - I(X_p ; X_r)
    \leq H(X_r)
    \label{equation:bound}
    \\
    &= H(X|X_p) + I(X_p; X_r)
    \label{equation:residual}
    .
\end{align}
Here \eqref{equation:summands} uses the fact that having observed $X_p$, the only uncertainty in $X_r + X_p$ is due to $X_r$. The inequality in \eqref{equation:bound} uses the non-negativity of mutual information. $H(X_r)$ is rewritten in \eqref{equation:residual} using the definition of mutual information and once again the fact that given $X_p$, the only uncertainty in $X - X_p$ is due to $X$.

The term $I(X_p ; X_r)$ in \eqref{equation:bound} acts as a penalty term on the residual formulation. To minimize it, the residual $X_r$ and the prediction $X_p$ must be as independent from each other as possible. This can be achieved when $X_p$ collapses values in $X$ so that $H(X_r)$ decreases, or when $X_p$ produces a constant value for different values in $X$, reducing $H(X_p)$. Reducing $H(X_p)$ increases $H(X|X_p)$, which in turn could have the adverse effect of increasing $H(X_r)$, as shown in \eqref{equation:residual}.

In our proposed method we train to minimize $D_r$ and $\hat{H}(Y_r)$. By extension, we also minimize $\hat{H}(X_r)$. As shown in \eqref{equation:residual}, this reduces both $H(X|X_p)$ and $I(X_p; X_r)$. Hence, this optimization procedure encourages the learnable function $h_r(\cdot)$ to create a representation $X_p$ that recovers the input $X$ as accurately as possible, while at the same time being as independent as possible from the resulting residual $X_r$.

Note that enforcing the similarity of $X_p$ and the original input $X$ may not be an optimal procedure, since even though such an optimization will decrease $H(X|X_p)$, it may lead to an increase in $I(X_p ; X_r)$. This explains why in preliminary experiments, we found that having a function $h_r(\cdot)$ that explicitly reconstructs $X$ does not perform as well as our proposed method.

Due to the previous considerations stemming from our proposed method, we find that $H(Y_c|Y_t) = H(Y_r)$ can be easier to achieve. This motivates us to compare $\hat{H}(Y_b) + \hat{H}(Y_r)$ against the same baselines used in the conditional approach.

\subsection{Entropy modelling}
\begin{figure*}[t]
  \centering
  \begin{subfigure}{.49\textwidth}
      \centering
      \includegraphics[width=\textwidth]{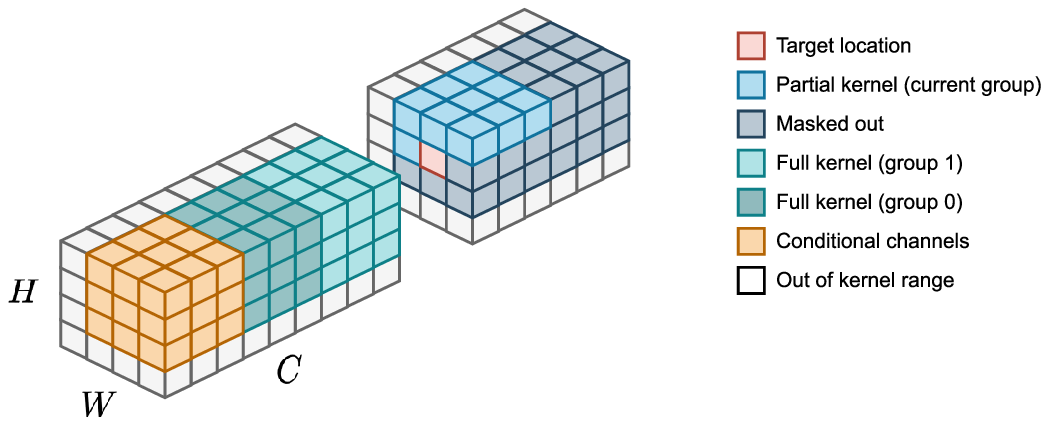}
      \caption{Convolution mask}
      \label{figure:entropy-mask}
  \end{subfigure}
  \begin{subfigure}{.49\textwidth}
      \centering
      \includegraphics[width=.65\textwidth]{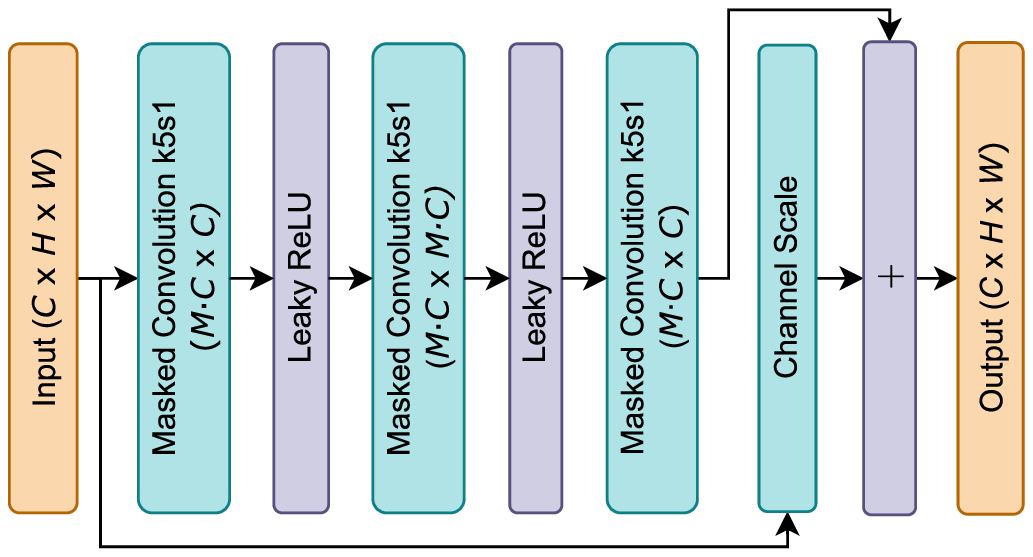}
      \caption{CNN block}
      \label{figure:entropy-block}
  \end{subfigure}
  \caption{Entropy model overview. (a) The convolution has kernel size $3 \times 3$ and the input is $12 \times 4 \times 4$. The conditional input has size $3 \times 4 \times 4$ and there are $K=4$ groups. With an input padding and a stride of 1, this is the 7th step of the convolution.}
\end{figure*}

To conditionally code a representation $Y_c$ that exploits as much information as possible from $Y_t$, we model the spatial and dimensional dependencies between and within representations using a CNN \cite{MinnenS20}. 
Our proposed entropy model strikes a balance between complexity and accuracy. We group channels with a fixed size $K$ \cite{MinnenS20}. Within each group, the same location across channels are processed in parallel, using as context all locations in the previous groups and all previous locations across all channels of the current group, within the receptive field of the convolutional layer. Similarly to \cite{balle2018variational}, locations in the spatial domain are processed in a top-to-bottom, left-to-right fashion. The Markov property is enforced by a mask applied to the convolution kernels. Fig.~\ref{figure:entropy-mask} shows the kernel mask for a single output channel of a layer.

Unlike previous work, the CNN architecture of our entropy model has scalable residual connections and deeper layers with kernels sizes larger than one for its auto-regressive convolutions. This removes some of the modelling limitations imposed by similar entropy models. The CNN architecture has blocks of three layers in which the input channels are scaled up, transformed in a higher-dimensional space, and scaled down back to the original number of channels. The residual connections are introduced in-between these blocks, such that the inputs can be re-scaled differently across the channel dimension. To maintain the Markov property when the number of channels changes, the group sizes are re-scaled accordingly and the channels can only change in multiples of $M$. Fig.~\ref{figure:entropy-block} shows the architecture overview of a single block in the CNN.

The conditional representation is available as another channel group and is transformed by the CNN in the same way as other groups, except that its context is restricted to the receptive field within that group. All elements of the conditioned representation have access to this group.

Similarly to \cite{balle2018variational} and more recent works, the predictions of the entropy model correspond to the means $W$ and scales $\Sigma$ of a univariate Gaussian distribution assigned to each element in the representation. The symbols are obtained by $Q = \lfloor Y - W \rceil$, and the corresponding probability is $P_{\mathcal{N}(\mathbf{0}, \Sigma)}[\vert Q \vert \leq \vert Q \vert \pm \nicefrac{1}{2}]$. During training, the rounding operation is simulated by adding uniform noise $\mathcal{U}(-\nicefrac{1}{2}, \nicefrac{1}{2})$.

\subsection{Learnable scalable compression}

As an architecture for learnable compression, we use a simplified version of the work in \cite{HeYPMQW22}. We drop the side information components from the coder, introduced as a hyper-prior in \cite{balle2018variational}. We also remove the attention layers introduced in \cite{Cheng2020LearnedIC}. To reduce the memory footprint and speed up the training procedure, we incrementally reduce the channels in the first layers of the analyzers and the last layers of the synthesizers.

The base and enhancement tasks have this same architecture. The base generates a representation with the same dimensionality and resolution as the input $X$ but reconstruction of the input is not enforced. This representation is the input for the computer vision model. The coder and the computer vision model are trained together end-to-end.

In the conditional approach, $h_c(\cdot)$ is a traditional CNN composed of blocks with residual connections between them. Each block has three convolutional layers that perform a transformation in a higher dimensionality and scales the output back to a lower dimensionality. The first half of the blocks maintain the same dimensionality as $Y_b$, while the second half transitions to the same dimensionality as $Y_c$, obtaining $Y_t$. The resolution is maintained across the network as both $Y_b$ and $Y_c$ have the same resolution. In the residual approach, $h_r(\cdot)$ uses the same architecture as our synthetizers, to transform $Y_b$ into $X_p$. The synthetizer upscales the representation to match the resolution and dimensionality of $X$.

A representation $Y_b$ that is optimized for $D_b$ can contain information that might not be beneficial for reconstruction on its own. Moreover, the information in $Y_b$ is represented in a way that is suitable for the computer vision task $T$ and bringing it all back to the image feature space through $h_r(\cdot)$ can be challenging. To overcome these obstacles, we add a small reconstruction penalty on a transformed $Y_b$ to the rate-distortion Lagrange minimization formulation:
\begin{align}
    \mathcal{L}_b = D_b + \lambda_b \hat{H}(Y_b) + \beta \, \mathbb{E}[d_e(\hat{h}_r(Y_b), X)]
    ,
\end{align}
where $\hat{h}_r(\cdot)$ is an auxiliary network with the same architecture as the other syntherizers and $\lambda_b$ and $\beta$ are hyper-parameters. For the enhancement representations, we use the traditional rate-distortion loss function \cite{tishby2000information}:
\begin{align*}
    \mathcal{L}_c = D_c + \lambda_e \hat{H}(Y_c | Y_t), &&
    \mathcal{L}_r = D_r + \lambda_r \hat{H}(Y_r),
\end{align*}
for the conditional and residual approaches, respectively. During training, either the base network remains frozen or the gradients from the reconstruction network do not flow into the base network.

\section{Experiments}
\begin{figure*}[t]
  \centering
  \begin{subfigure}{.49\textwidth}
      \centering
      \includegraphics[width=\textwidth]{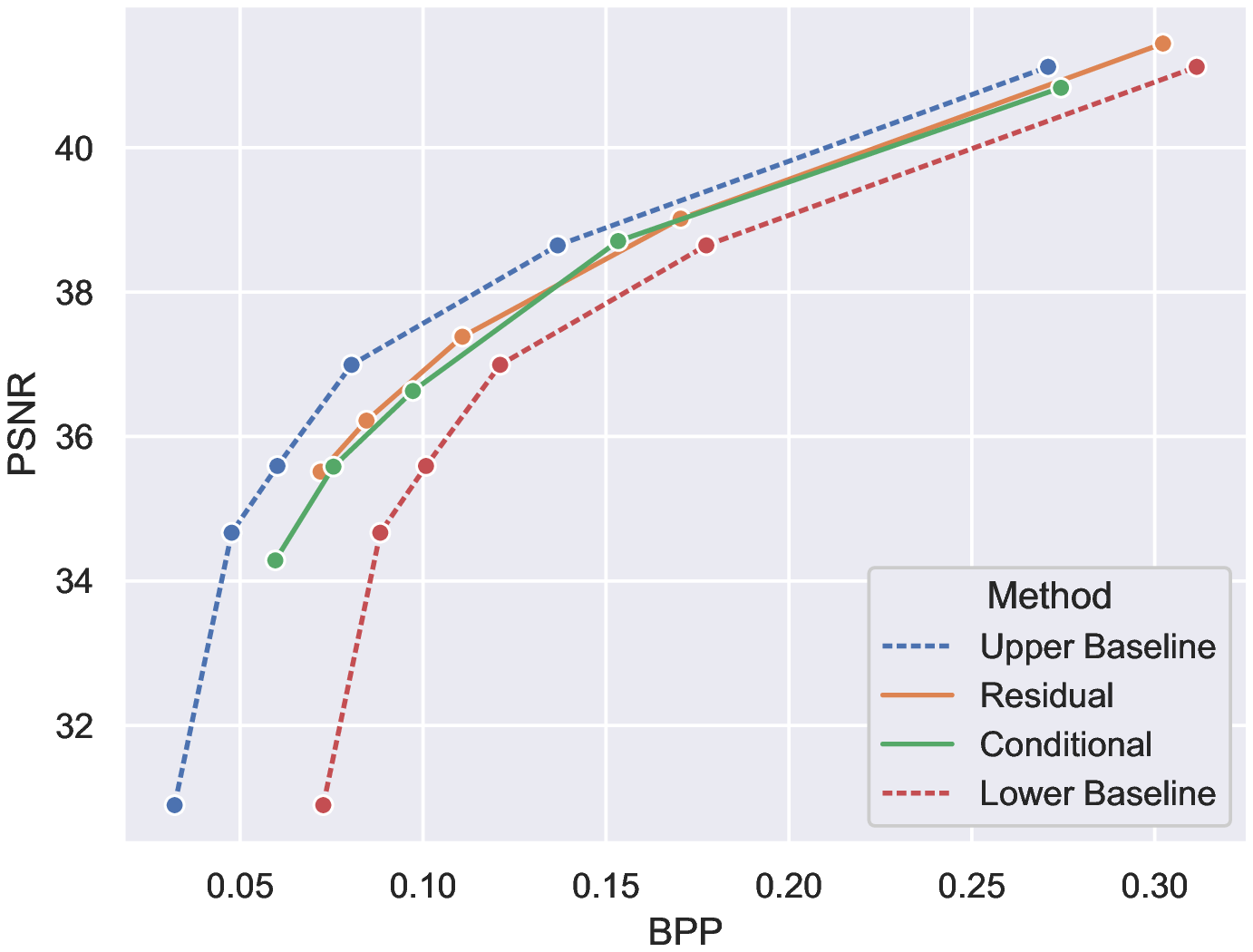}
      \caption{Scalable coding rate-distortion curves for Cityscapes}
      \label{figure:segmentation-rd}
  \end{subfigure}
    \begin{subfigure}{.49\textwidth}
      \centering
      \includegraphics[width=\textwidth]{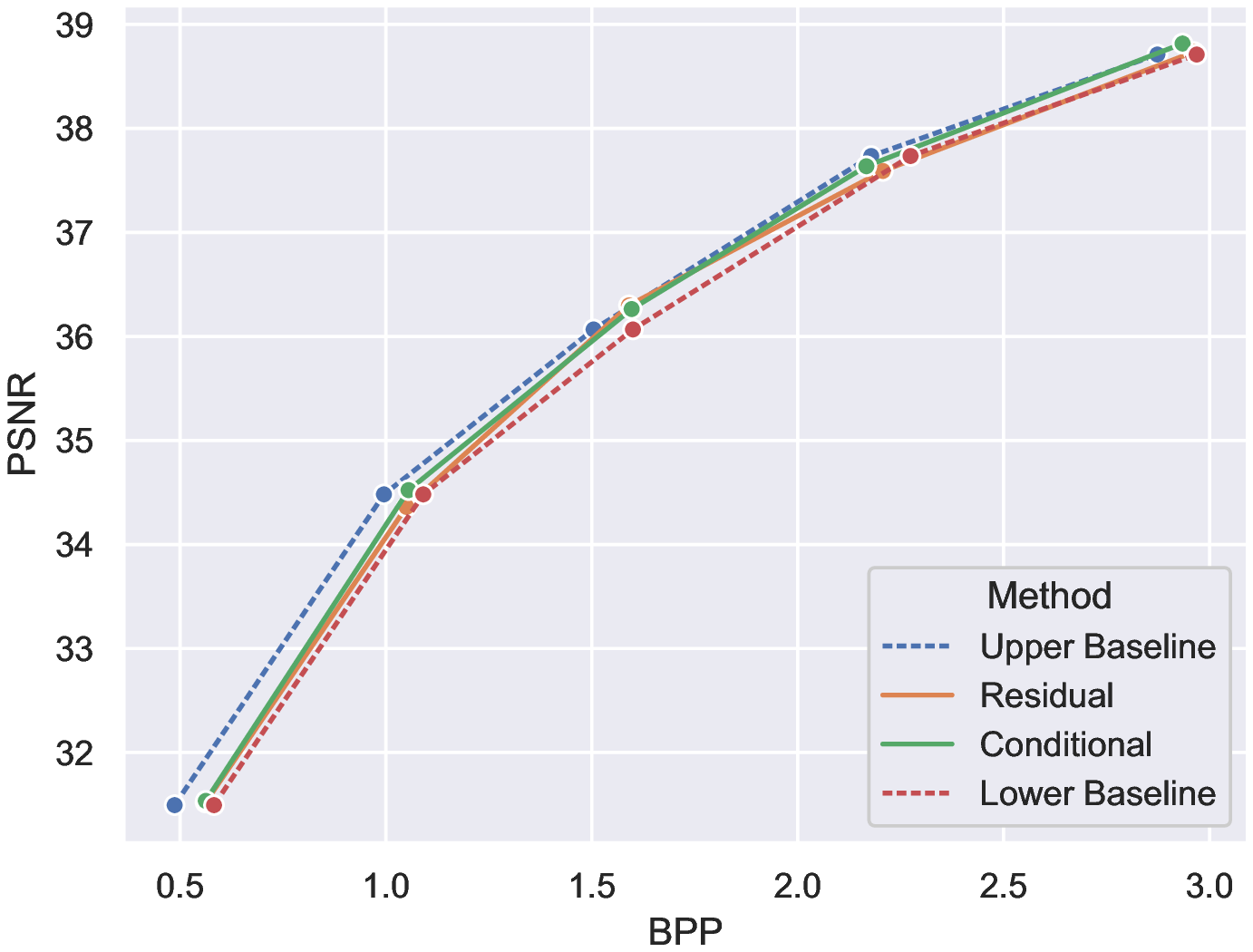}
      \caption{Scalable coding rate-distortion curves for COCO}
      \label{figure:detection-rd}
  \end{subfigure}
  \begin{subfigure}{.49\textwidth}
      \centering
      \includegraphics[width=\textwidth]{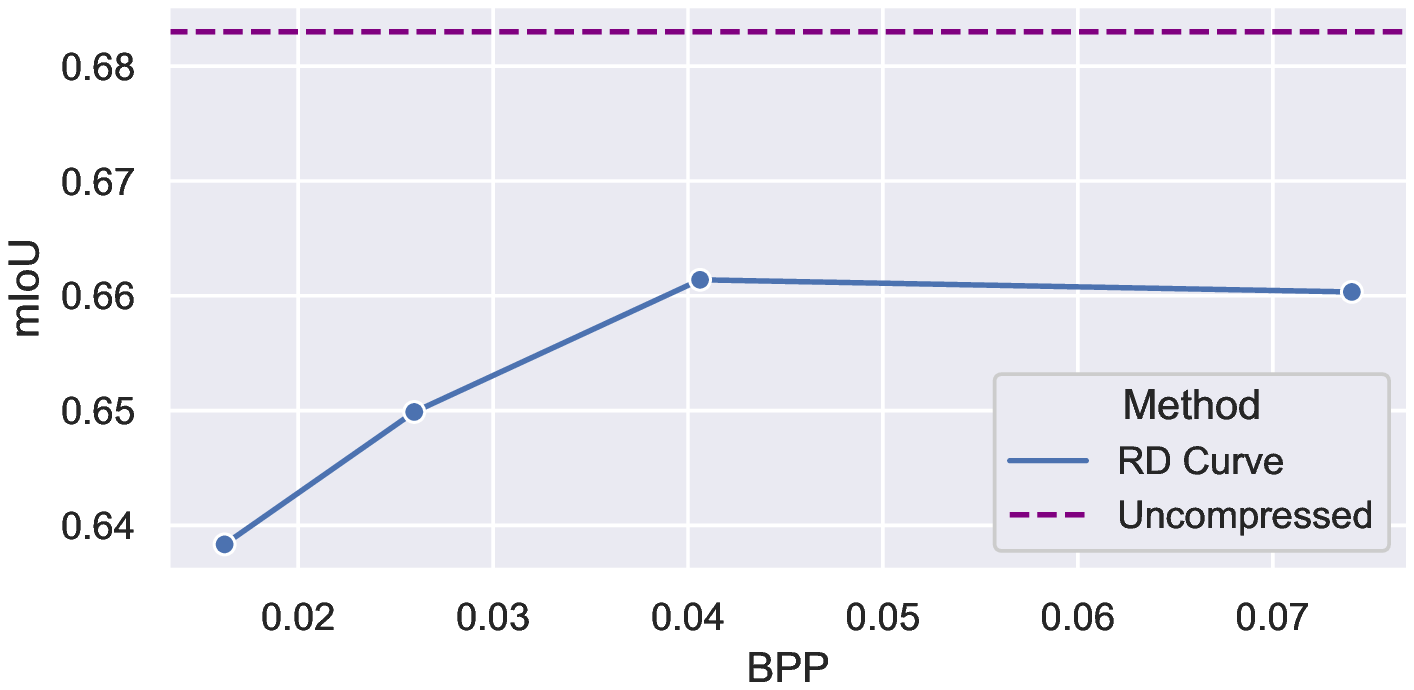}
      \caption{Rate-distortion curve for semantic segmentation on Cityscapes}
      \label{figure:segmentation-base}
  \end{subfigure}
    \begin{subfigure}{.49\textwidth}
      \centering
      \includegraphics[width=\textwidth]{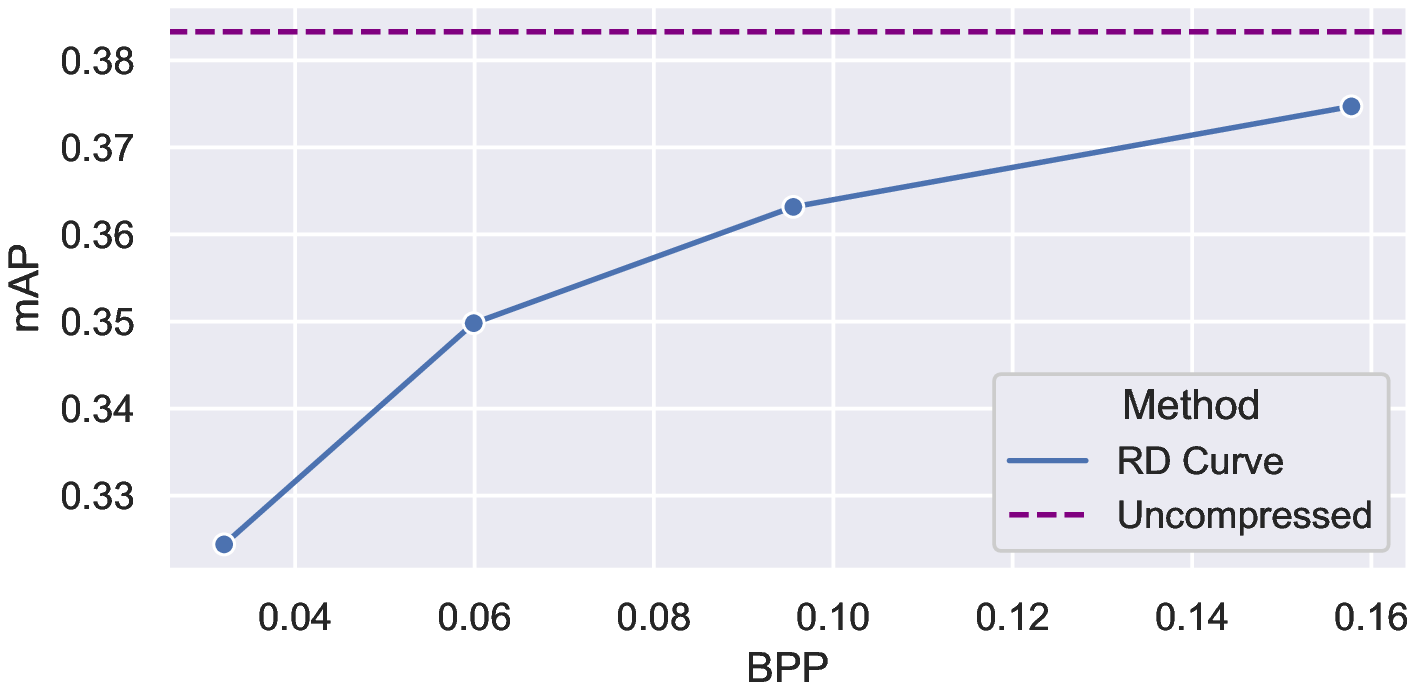}
      \caption{Rate-distortion curve for object detection on COCO}
      \label{figure:detection-base}
  \end{subfigure}
  \caption{Scalable coding results. The purple lines represent the performance attained with $\lambda_b = 0$ and $\beta = 0$.}
\end{figure*}

We conduct experiments to analyze the rate-distortion performance of the proposed conditional and residual methods for scalable coding and compare them against the proposed baselines. 
We perform two sets of experiments: 
one using 
semantic segmentation as the computer vision task on the Cityscapes dataset~\cite{cordts2016cityscapes}, and another 
using object detection as the computer vision task on the COCO 2017 dataset~\cite{LinMBHPRDZ14}.

We first train the base representation on the computer vision task to obtain $f_b(\cdot)$ and $g_b(\cdot)$ for rate-distortion points under different values of $\lambda_b$. We choose a model corresponding to a point on the rate-distortion curve that achieves reasonable distortion and subsequently use it to generate the representations $Y_b$ for the conditional and residual approaches. The upper baseline is created by training the reconstruction task with no side information. We use the same architecture for the analysis and synthesis transforms as the other models. The entropy model for the upper baseline has the same architecture as the one used in the residual approach. The lower baseline is obtained by adding the rate of the base representation used for the conditional and residual approaches.

Across all experiments, we allocated $C_b = 32$ channels to the base representation and $C_e = 256$ channels to the enhancement representation. In the analysis transforms, the four down-scaling operations have output channels 24, 48, 192, and $C_b$ or $C_e$, while the up-scaling operations in the synthesis transform have output channels 192, 48, 24, and $C_x$. The entropy model consists of 5 blocks with $K = 16$ and $M = 1$.

To train the reconstruction tasks in all experiments, we use the RMSE function as the distortion function $d_e(\cdot,\cdot)$. We compute and report the bits-per-pixel (BPP) using the entropy estimates, which in several experiments had at most 0.5\% difference with the achieved BPP. Also, to speed up the computation of the rate-distortion curve, we often train a model under low-compression settings and use its weights as initialization for the models trained to obtain the rest of the curve. The parameters are updated using Adam at a learning rate of $10^{-4}$. We train models with early stopping but first decay the learning rate by a factor of 0.75 if a plateau is reached.

\subsection{Image semantic segmentation on Cityscapes}

Cityscapes is a set of images of urban scenes for semantic understanding \cite{cordts2016cityscapes}. We use DeepLabV3 \cite{ChenPSA17} as the computer vision model for segementation with MobileNetV3 \cite{HowardPALSCWCTC19} as a back-end. Here, $d_b(\cdot, \cdot)$ is the per-pixel multi-class cross-entropy, although we report the mean intersection over union (mIoU) metric. We set $\beta = 0.1$ and report the results on the validation dataset. For data augmentation, we use random crops of $768 \times 768$ pixels, random horizontal flips and color jittering. The front-end of the model corresponding to the coder is trained with Adam using a learning rate of $10^{-4}$, while the classifier is trained with stochastic gradient descent using momentum and a learning rate of $10^{-2}$. A $\ell_2$ loss is added to the weights of the classifier to prevent over-fitting, with a scale factor of $10^{-4}$. 

Fig.~\ref{figure:segmentation-rd} shows the rate-distortion curves for the conditional and residual approach. We notice that these lines lie in between the baselines, producing rate-distortion points that respect them. Compared to the rate-distortion curve of the lower baseline, the conditional approach has a BD-Rate \cite{bjontegaard2001calculation} of $-16.56\%$, whereas the residual approach achieves a $-14.6\%$ rate reduction. Thus, the conditional approach performs marginally better than the residual approach in terms of BD-Rate. Looking at the ratio between these BD-Rate scores and the BD-Rate score achieved by the upper baseline, we can compute the percentage of the base representation utilized. As such, the conditional approach uses $43.01\%$ of the side information rate, whereas the residual approach uses $37.91\%$. In the lowest-compression settings under both approaches, the utilization is higher.

Fig.~\ref{figure:segmentation-base} shows the rate-distortion performance of the base task. The chosen $\beta$ value places a penalty on both the rate and the task performance but allows the base representation to be exploited by the architecture. We attribute the small imperfections in this rate-distortion curve to the choice of $\beta$ and the limitations of the training algorithm.

\subsection{Object detection on COCO}
COCO 2017 has 123,287 domain-agnostic images for object detection and segmentation \cite{LinMBHPRDZ14}. We use Faster R-CNN \cite{RenHG017} for object detection with ResNet-50 \cite{HeZRS16} as the back-end. For this task, $d_b(\cdot, \cdot)$ is the sum of the different loss functions employed by this architecture. We report the mean average precision (mAP) metric computed according to \cite{LinMBHPRDZ14}, and set $\beta = 0.05$. As data preprocessing for training, we use random horizontal flips and generate batches with similar aspect ratios grouped in 3 clusters. Images inside a batch are resized to the minimum size and the bounding boxes are adjusted accordingly when training the computer vision task. When training for reconstruction, the images in a batch are center-cropped to their minimum size. All weights are trained with Adam using a learning rate of $10^{-4}$.

As shown in Fig.~\ref{figure:detection-rd}, the performance of both approaches is comparable, with a $-4.14\%$ and a $-2.47\%$ BD-Rate improvement over the lower baseline, for the conditional and residual methods, respectively. We achieve a utilization of the base in terms of BD-Rate of 49.24\% and 29.32\% for the conditional and residual approaches, respectively. Fig.~\ref{figure:detection-base} shows the base-distortion performance of the base task. Compared to semantic segmentation on Cityscapes, the task model is better at reaching the uncompressed task performance. Also, for a similar distortion penalty, this task uses more rate. The rate-distortion curves obtained by the reconstruction task on the COCO dataset are almost an order of magnitude larger than the ones from Cityscapes. This can be explained by the simplicity of the content in the images found in Cityscapes, and the higher amount of artifacts found in the COCO dataset due to compression.

\section{Conclusion}
We present conditional and residual methods for scalable coding for humans and machines. Our experiments show that the proposed architectures for conditional and residual coding perform similarly and that the rate-distortion performance is within the presented baselines or operational bounds. In addition, the proposed conditional entropy model is able to match the performance of the residual method.


\bibliographystyle{IEEEbib}
\bibliography{main}

\end{document}